\documentclass{dcds-b}
\usepackage{algorithmic}
\usepackage{amssymb}  
\usepackage{algorithm}
\usepackage{multirow}
\usepackage[tight,footnotesize]{subfigure}
\usepackage{amsmath}
  \usepackage{paralist}
  \usepackage{graphics} 
  \usepackage{epsfig} 
\usepackage{graphicx}  \usepackage{epstopdf}
 \usepackage[colorlinks=true]{hyperref}
\hypersetup{urlcolor=blue, citecolor=red}

\def\currentvolume{19}
 \def\currentissue{5}
  \def\currentyear{2014}
   \def\currentmonth{July}
    \def\ppages{1335--1354}
     \def\DOI{10.3934/dcdsb.2014.19.1335}

\newcommand{\BEQ}{\begin{equation}}
\newcommand{\EEQ}{\end{equation}}
\newcommand{\BEA}{\begin{eqnarray}}
\newcommand{\EEA}{\end{eqnarray}}

\newcommand{\bx}{{\bf x}}

\newcommand{\bp}{{\bf p}}

\renewcommand{\H}{{\mathcal H} }
\renewcommand{\L}{{\cal L} }
\newcommand{\ssum}{{\sum}}

\newcommand\Mark[1]{\textsuperscript#1}

\theoremstyle{definition}

\title[Latent Self-Exciting Point Process Model]
      {Latent Self-Exciting Point Process Model for \\Spatial-Temporal Networks}

\author[Yoon-Sik Cho, Aram Galstyan, P. Jeff Brantingham and George Tita]{}


 \email{yoonsik@isi.edu}
 \email{galstyan@isi.edu}
 \email{branting@ucla.edu}
 \email{gtita@uci.edu}

\begin{document}
\maketitle

\centerline{\scshape Yoon-Sik Cho}
\medskip
{\footnotesize
 \centerline{USC Information Sciences Institute}
   \centerline{Marina del Rey, CA 90292, USA}
} 

\medskip

{\centerline{\scshape  Aram Galstyan\Mark{1}, P. Jeff Brantingham\Mark{2} and George Tita\Mark{3}}}
\medskip

{\footnotesize
 \centerline{\Mark{1}USC Information Sciences Institute, USA}
 \centerline{\Mark{2}University of California, Los Angeles, USA}
   \centerline{\Mark{3}University of California, Irvine, USA}
}

%

\begin{abstract}
We  propose a latent self-exciting point process model that describes geographically distributed interactions between pairs of entities. In contrast to most existing approaches that assume fully observable interactions, here we consider a scenario where certain interaction events lack information about participants. Instead, this information needs to be inferred from the available observations. We develop an efficient approximate algorithm based on variational expectation-maximization to infer unknown participants in an event given the location and the time of the event. We validate the model on synthetic as well as real-world data, and obtain very promising results on the identity-inference task. We also use our model to predict the timing and participants of future events, and demonstrate that it compares favorably with baseline approaches.
\end{abstract}

\section{Introduction}

In recent years there has been a considerable interest in understanding dynamic social networks.  Traditionally, longitudinal analysis of social network data has been limited to relatively small amounts of data collected from manual and time-consuming surveys. Recent development of various sensing technologies, online communication services, and  location-based social networks has made it possible to gather time-stamped and geo-coded data on social interactions at an unprecedented scale. Such data can potentially facilitate better and more nuanced understanding of geo-temporal patterns in social interactions. To harness this potential, it is imperative to have efficient computational models that can deal with spatial-temporal social networks.

One of the main challenges in social network analysis is  handling missing data. Indeed, most social network data are generally incomplete, with missing information about  links~\cite{Guimera2009,Kossinets2006,Kleinberg2007}, nodes~\cite{eyal:identifying} or  both~\cite{KimLescovec}. In repeated interaction networks studied here, there is another source of data ambiguity that comes from limited observability of certain interaction events. Namely, even when interactions are recorded, information about participants might be missing or only partially known. A real-world problem that highlights the latter scenario is concerned with inter-gang rivalry network in Los Angeles, where the records of violent events between rival gangs might lack information about one or both participants~\cite{OT}.

\begin{figure}[!t]
\vskip 0.2in
\centering
\includegraphics[width=0.65\columnwidth]{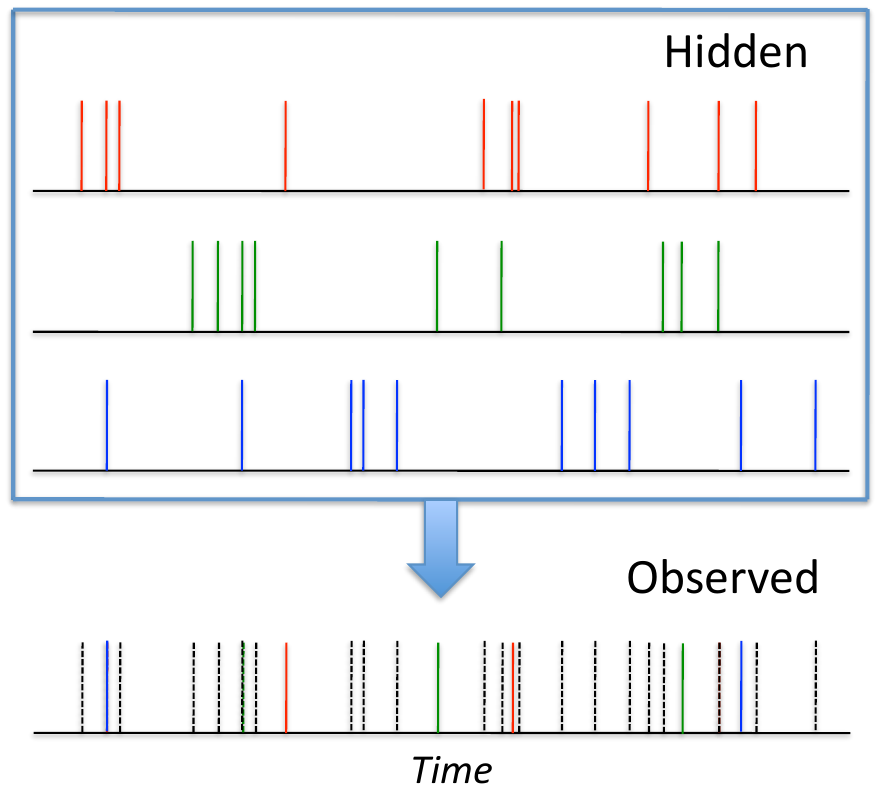}
\caption{Schematic demonstration of the missing label problem for temporal point processes. The dashed lines represent events for which the generating process is unknown.}
\label{fig:tworelation}
\end{figure}

Here we  formalize the missing label problem for spatial-temporal  networks by introducing Latent Point Process Model, or LPPM, to describe geographically distributed interaction events between pairs of entities. LPPM assumes that interaction between each pair is governed by a spatial-temporal point process. In contrast to existing  models~\cite{NIPS2011_1106,pointprocess,PoissonNetwork,PoissonCascade}, however, it allows a non-trivial generalization where certain attributes of those events are not fully observed. Instead, they need to be inferred from available observations. To illustrate the problem, consider  a sequence of events generated by $M$ temporal point processes; see Figure~\ref{fig:tworelation}. Each sequence is generated via a non-homogenous  and  possibly history-dependent point process.  The combined time series is a {\em marked} point process, where the mark, or the label, describes the component that generates the event. The observed data consists of the recorded events. However we assume that those labels are only partially observable, and need to be inferred from the observations.

How well can one identify the label of a specific event based on limited observations? The answer depends on the nature of the process generating the events. For instance, if the events in Figure~\ref{fig:tworelation} are generated by a set of independent and homogenous Poisson processes with intensities $\lambda_1< \lambda_2< \lambda_3$,  then the identification accuracy is limited by $\frac{\lambda_3}{\lambda_1+\lambda_2+\lambda_3}$, i.e., all the unlabeled events are attributed to the process with the highest intensity. Luckily, most real-world processes describing human interactions demonstrate highly non-homogenous and history-dependent temporal patterns, suggesting that interaction events are not statistically independent,  but exhibit  non-trivial correlations~\cite{Barabasi2005,PhysRevE.81.035101}. To account for temporal correlations,  here we augment LPPM with a  model of  {\em self-exciting point process} known as Hawkes process, which has been used previously in a number of applications. Furthermore, we use interaction-specific mixture distributions of spatial patterns of interactions to inform the inference problem.

Learning and inference with LPPM constitutes inferring missing labels, predicting the timing and/or the source of the next event, and so on. Due to missing observations, exact inference and learning is intractable for even moderately large datasets. Toward this end,  we develop an efficient algorithm for learning and inference based on the variational EM approach~\cite{Bernardo03thevariational}. We validate our model for both synthetic and real-world data. For the latter, we use two distinctly different datasets (1) data on inter-gang violence from Los Angeles Police Department; (2) User check-in data from Gowalla, which is a location based social networking service. Our results indicate that LPPM  is better than baselines in both inference and prediction tasks.

The rest of the paper is organized as follows: After reviewing some relevant work in Section~\ref{sec:related}, we define our latent point process model in Section~\ref{sec:model}. In Section~\ref{sec:inference}  we describe variational EM approach for efficient learning and inference with LPPM. We present results of experiments with both synthetic and real-world data in Sec.~\ref{sec:results-synthetic} and~\ref{sec:real-world}, and provide some concluding remarks in Section~\ref{sec:conclusion}.

\section{Related work}
\label{sec:related}
Modeling temporal social networks has attracted considerable interest in recent years. Both discrete-time~\cite{Hanneke2009} and continuous-time~\cite{Fan2009,Steglich07interuniversitycenter,VuAHS11} models have been proposed to study longitudinal networks. In particular, Perry and Wolfe~\cite{pointprocess} suggested point process models for describing repeated interactions among a set of nodes. They used a Cox hazard model and allowed the interaction intensity  to depend on the history of interactions as well as on node attributes. In contrast to our work, however, Ref.~\cite{pointprocess}  assumes that all the interactions are perfectly observable. Different continuos time models such as  Poisson Cascades~\cite{PoissonCascade}, Poisson Networks~\cite{PoissonNetwork}, and Piecewise-Constant Conditional Intensity Model~\cite{NIPS2011_1106}, have also been used to describe temporal dependencies between events.

In a related line of research, a number of authors have addressed the problem of uncovering hidden networks that facilitate information diffusion and/or activation cascades,  based on  time-resolved traces of such cascades. 
  Most of the existing approaches rely on temporal information only~\cite{GomezKDD,InfoPath,survival,du,versteeg2012}, although several other methods also utilize additional features, such as prior knowledge about the network structure~\cite{Netrapalli}, or the content  diffusing  through the network~\cite{wang,versteeg2013}.

Self-exciting point process was originally suggested in seismology  to model aftershock of earthquakes~\cite{Seismology}. Self-exciting models have since been used in a number of diverse applications, such as assessing financial portfolio credit risk~\cite{portfolio}, detecting terrorist activities~\cite{Terror}, predicting corporate defaults~\cite{Azizpour08self-excitingcorporate}. Recently, Mohler et al.~\cite{Mohler:2011:SEP} used spatial-temporal self-exciting process to model urban crime. Their model, however, studies a different problem and does not  assume any  missing information. In particular, they consider a univariate point process, as opposed to multi-variate model used here, which is needed  to describe  interactions among different entities.

Stomakhin et al.~\cite{OT} studied the temporal data reconstruction problem in a very similar settings. Their approach, however,  assumes known model parameters, which is impractical in real-world scenario, thus limiting their experiments to synthetically generated data only. In contrast, LPPM learns  the model parameters directly from the data labeled or unlabeled. More recently, Hegemann et al.~\cite{Hegemann} proposed a method that does not assume known parameters but learns those parameters using an iterative scheme.  The main difference between LPPM and Ref.~\cite{Hegemann} is that  the former is a generative probabilistic model, which allows to estimate the {\em posterior probability} that a certain pair is involved in a given event based on the observations. Ref. \cite{Hegemann}, on the other hand, calculates heuristic  score-functions that can be used to rank different pairs'  involvement as more or less likely, but those scores cannot be interpreted as probabilities. Furthermore,  in contrast to Ref.~\cite{Hegemann} , here we consider both temporal and spatial components, and use the generative model for event prediction.

\section{Spatial-temporal model of relationship network}
\label{sec:model}
Consider $N$ individuals forming $M$ pairs that are engaged in pairwise interactions with each other. Generally $M$ would be the total number of undirected edges that $N$ has, which is $N(N-1)/2$. However, for some cases (i.e., the network structure is given or some pairs are out of our consideration) total number of pairs $M$ could be fixed to the size of our interest for efficient computation. We  observe a sequence of interaction events (called events hereafter) given as $\H=\{h_k\}_{k=1}^n$, where each event is a tuple $h_{k}=(t_k,\bx_k, z_k)$. Here  $t_k\in \mathbb R^+$ and $\bx_k \in \mathbb R^2$ are the time and the location of the event, while $z_k$ is the symmetric interaction matrix for the event $k$: $z_k^{ij}=1$ the agents $i$ and $j$ are involved in the $k$-th event, and $z_k^{ij}=0$ otherwise. Since each event involves only one pair of agents, we have $\sum_{i<j}z_k^{ij}=1$. Without loss  of generality, we assume $t_1=0$ and $t_n=T$.

Let $\H_t$ denote the history of events up to time $t$ as the set of all the events that have occurred before that time, $\H_t= \{ h_k\}_{t_k<t}$.
 We assume that the interactions between the pairs are point processes with spatial-temporal conditional  intensity function $S_{ij}(t, \bx|\H_t)$, so that the probability that the agents $i$ and $j$ will interact within a time window $(t,t+d)]$ and location $(\bx,\bx+d\bx)$ is simply $ S_{ij}(t, \bx|\H_t)dt d \bx$. Note that the intensity function is conditioned on the history of past events. Here we assume that the above intensity function can be factorized into temporal and spatial components as follows:
\BEQ
S_{ij}(t, \bx|\H_t)=\lambda_{ij}(t|\H_t)r_{ij}(\bx)
\EEQ
The intensity function $S_{ij}(\cdot)$ is based on the separability of spatio-temporal covariance functions~\cite{cressie+w_11} assuming that the temporal evolution proceeds independently at each spatial location, but rather on the history of its own. Note that the temporal conditional  intensity $\lambda_{ij}(t|\H_t)$ is history-dependent, whereas the spatial component is not. The scope of our research is not the influence of spatial preference between nodes, but rather the spatial activities of pairs. In this regard, we assume that the pair's preference of location stays the same over time.
Let us define
\BEQ
\Lambda_{ij}^T = \int_{0}^{T} \lambda_{ij}(\tau|\H_\tau) d\tau
\label{eq:Lambda}
\EEQ

The likelihood of an observed sequence of interactions under the above model is given as
\begin{equation}
\bp(\H;\Theta)= \prod_{k}\prod_{i<j} \underbrace{ [ \lambda_{ij} (t_k)]^{z^{ij}_k} e^{-  \Lambda_{ij}^T    } }_{\text{temporal process}}  \underbrace{ [r_{ij} (\bx_k)  ]^{z^{ij}_k} }_{\text{spatial process}} \label{eq:likelihood}
\end{equation}
where the products are over all the events and the pairs, respectively. Here $\Theta$ encodes all the hyperparameters of the model (to be specified below).  From this point, we simplify the intensity expression to $\lambda_{ij}(t)$ omitting $\H_t$.

So far our description has been rather general. Next we have to specify a concrete parametric form of the temporal and spatial conditional intensity functions. As stated above, the consideration of non-trivial temporal correlations between the events suggest that it is not realistic to use Poisson process with constant intensity.  Instead, here we will use a Hawkes Process, which is a variant of a self-exciting process~\cite{Seismology}.

\subsection{Hawkes process}
We assume that the intensity of events involving the pair $(i,j)$ at time $t$ is given as follows:
 \begin {equation}
\lambda _{ij} (t) = \mu_{ij} + \sum _{p:t_p<t} g_{ij} (t-t_p)
\label{eq:1}
\end{equation}
where the summation in the second term is over all  the events that have happened up to time $t$. In Equation~\ref{eq:1}, $\mu_{ij}$ describes the background rate of event occurrence that is time-independent, whereas the second term describes the self-excitation part, so that the events in the past  increase the probability of observing another event in the (near) future.  We will use a two-parameter family for the self-excitation term:
\begin {equation}
\label{eqn:selfexciting}
g_{ij}(t-t_p) = \beta_{ij} \omega_{ij} \exp\{-\omega_{ij} (t-t_p)\}
\end{equation}
Here $\beta_{ij}$ describes the weight of the self-excitation term (compared to the background rate), while $\omega_{ij}$ describes the decay rate of the excitation.

\subsection{Spatial Gaussian Mixture Model (GMM)}
To model the spatial aspect of the interactions, we assume that different pairs might have different geo-profiles of interactions. Namely, we assume that the interaction of specific pair is spatially distributed according to a pair-specific Gaussian mixture model:
\begin{equation}\label{eq:GMM}
r_{ij}(\bx) = \ssum_{c=1}^C w_{ij}^c  {\mathcal N}(\bx; {\bf m}_{ij}^c,\Sigma_{ij}^c)
\end{equation}
 In Equation~\ref{eq:GMM}, $C$ is the number of components, $ {\mathcal N}(\bx; {\bf m}_{ij}^c,\Sigma_{ij}^c) $ denotes for 2-D multi-variate normal distribution with mean ${\bf m}_{ij}^c$ and covariance $\Sigma_{ij}^c$, and $w_{ij}^c$ is the weight of $c$-th component for pair $i,j$. The number of components $C$ was obtained using Ref.~\cite{Pelleg:2000:XEK:645529.657808}, where BIC scores were used to optimize the number of components. More weights on specific component on space means more chances of appearance within the cluster of the component. For simplicity, the dynamics of the weights over time has been ignored.

We would like to note that the use of Gaussian mixtures rather than a single Gaussian model is justified by the observation that interactions among the same pair might have different modalities (e.g., school, or movies, etc.). In this sense, the model borrows from the mixed membership stochastic block model~\cite{MMSB},  which assumes that the agents can interact while assuming different roles.

%

Equations~\ref{eq:Lambda}-\ref{eq:GMM} complete the definition of our latent point process model. Next we describe our approach for efficient learning and inference with LPPM.

\section{Learning and inference}
\label{sec:inference}
As mentioned in the introduction, we are interested in scenario where the actual participants of the events are not observed directly, and need to be inferred, together with the model parameters (i.e., pair-specific parameters of the Hawkes process model and the Gaussian mixture model). For the latter, we employ maximum likelihood (ML) estimation. ML selects the parameters that maximize the likelihood of observations, which consist of the timing and the location of the events, and participant information for some of the events.

Due to the missing labels, some of the interaction matrices $z_{k}$ are unobserved (or {\em latent}) for some $k$. Therefore, there is no closed-form expression for the likelihood of the observed sequence of events. Instead, one has to resort to approximate techniques for learning and inference, which is described next.  Here we use a variational EM approach~\cite{Bernardo03thevariational} by positing a simpler distribution $Q(Z)$ over the latent variables with free parameters. The free parameters are selected to minimize  the Kullback-Leibler (KL) divergence between the variational  and the true posterior distributions. Recall that the KL divergence between two distribution $Q$ and $P$ is defined  as
\begin{equation}
 D_{KL}(Q||P) = \int_Z Q(Z) \log {Q(Z)\over P(Z,Y)} dZ
 \end{equation}
 where $Z$ is the hidden variables, and $Y$ is the observed variables. In our case, $Z$ is the hidden identity of interaction where some of the portion is known, whereas $Y$ describes the location and the time of the incident.

 We introduce the following variational multinomial distribution:
 \begin{equation}
 Q(\mathcal{Z} _n  |  \Phi) = \prod_k \prod_{i<j} q(z_{k}^{ij} | {\phi}_k )
 \label{Q:var}
 \end{equation}
where $\mathcal{Z}_k=\{z_l\}_{l=1}^k$ denotes the set of interaction matrices for events up to the $k$-th event,  and $q(\cdot|\phi_k)$ being the multinomial distribution with parameter $\phi_k$. The matrix $\phi_k$ consists of the free variational parameters ${\phi}_{k}^{ij}$ describing the probability that the agents $i$ and $j$ are involved in the  $k$-th event. Note that the present choice of the variational distribution discards correlations between past and future incidents, thus making the calculation tractable.

The variational parameters are determined by maximizing  the following lower bound for the log-likelihood~\cite{Bernardo03thevariational}:
 \begin{eqnarray}
 \mathcal{L}_{\Phi} (Q,\Theta) &=& E_Q\bigl[ \log \prod_{k} \prod_{i<j} [\lambda_{ij} (t_k) ]^{z^{ij}_k} e^{-\Lambda_{ij}^T } \bigr] \nonumber \\
 &+& E_Q\bigl[\log \prod_{k} \prod_{i<j}  [r_{ij} (\bx_k) ]^{z^{ij}_k}  \bigr]  \nonumber \\
&-& E_Q[\log \prod _{k} \prod_{i<j} q(z_{k}^{ij} | {\phi}_k ) ]
 \label{eq:lowerbound}
 \end{eqnarray}
where $\Phi$  is the set of variational parameters, and $\Theta$ is the set of all the model parameters. The above equation can be rewritten as follows:
 \begin{eqnarray}
 \mathcal{L}_{\Phi} (Q,\Theta) &=& E_Q\bigl[ \sum_k \sum_{i<j} z_k^{ij} \log  [\lambda_{ij} (t_k) ] {-\Lambda_{ij}^T } \bigr] \nonumber \\
 &+&  \sum_k \sum_{i<j} {\phi^{ij}_k}  \log  [r_{ij} (\bx_k) ]   \nonumber \\
&-& \sum_k \sum_{i<j} \phi_k^{ij} \log   {\phi}_k^{ij}
 \label{eq:lowerbound2}
 \end{eqnarray}
 where in the last two terms we have explicitly performed the averaging over the multinomial variational distribution defined in Equation~\ref{Q:var}.

Variational EM algorithm works by iterating between the E--step of calculating the expectation value using the variational distribution, and the M--step of updating the model (hyper)parameters so that the data likelihood is locally maximized. The overall pseudo-algorithm is shown in Algorithm~\ref{alg:VEM}. The details of update equations used in both E--step and M--step are provided in the appendix.

\begin{algorithm}[tb]
   \caption{Variational EM}
   \label{alg:VEM}
\begin{algorithmic}
   \STATE{\bfseries Size:} consider total of $n$ events,  $M=\frac{N(N-1)}{2}$ pairs
   \STATE {\bfseries Input:} data $\bx_{1:n}$,  $t_{1:n}$,  $z_k$ of complete events
   \STATE Start with initial guess for hyper parameters.
   \STATE Fix all ${\phi}_{k} = z_k$ for labeled events.
   \REPEAT
    \STATE Initialize all components of ${\phi}_{k}$ corresponding to unknown pairs or event $k$ to $\frac{1}{M}$
   \REPEAT
   \FOR{$k=1$ {\bfseries to} $n$}
	\IF {the pairs of $k$-th event is unknown}
                \STATE Update  ${\phi}_{k}$ using Eq.~\ref{eq:phi_func}
        \ENDIF
   \ENDFOR
   \UNTIL{convergence across all time steps}

   \STATE Update hyper parameters.
   \UNTIL{convergence in hyper parameters}
\end{algorithmic}
\end{algorithm}
%
%

\section{Experiments with synthetic data}
\label{sec:results-synthetic}
We first report our experiments with synthetically generated data for six pairs of agents. The sequence of interaction events was generated according to the LPPM process as follows:
\begin{enumerate}
\item For each pair, sample the first time of the incident using an exponential distribution with rate parameter $\mu$.
\item For each pair, sample the duration of time until the next incident using Poisson thinning. Since we are dealing with non-homogeneous Poisson process, we use the so called thinning algorithm \cite{Thinning} to sample the next time of the event. By repeating step 2, we obtain the timestamps of incidents for each pair.
\item For every timestamp of a given pair we sample the location of the incident.
\end{enumerate}
To compare the performance of our algorithm with previous approaches, we follow the experimental set-up proposed in~\cite{OT}, where the authors used temporal-only information for reconstructing missing information in synthetically generated data. In addition to ML estimation, Ref.~\cite{OT} also used an alternative objective function  over relaxed continuous variables, and performed constrained optimization of the new objective function using $l^2$ regularization. Although  their method does not assign proper probabilities to the various timelines, it can provide a ranking of  most likely  participants.

\begin{table}[!h]
\caption{Model evaluation for total of n = 40 events between 6 pairs. Only 4 events have unknown participants.  The parameters are $\mu=10^{-2} \text{days}^{-1} $, $\omega=10^{-1} \text{days}^{-1} $ and $\beta = 0.5$. The accuracy of top three method is from Ref.~\cite{OT}, and Variational EM is our result using LPPM. The results are averaged over 1000 trials.}
\label{table:methodology_comparison}
\vskip 0.15in
\begin{center}
\begin{small}
\begin{sc}
\begin{tabular}{lcccr}
\hline
  method & Accuracy  \\
\hline
 Exact ML  & 47.3 \% \\
  max $l^1$  & 47\% \\
  max $l^2$  & 47.1\% \\
  Variational EM  & 46.9\% \\
\hline
\end{tabular}
\end{sc}
\end{small}
\end{center}
\vskip -0.1in
\end{table}
Following Ref.~\cite{OT}, we consider $40$ events, and assume that for $10\%$ (4 events) we do not have participant information. Table \ref{table:methodology_comparison} shows the overall performance of different approaches. To make the comparison meaningful, we omit the spatial information in our model, and focus on the temporal part only. For our algorithm the results are averaged over $1000$ runs.

Throughout this paper we measure the accuracy (expressed as a percentage) by counting the number of correctly identified events divided by the total number of hidden events. Table~\ref{table:methodology_comparison} indicates that  all four methods perform almost identically. In particular, all four methods have significantly better accuracy than the simple baseline value $1/6$, where each pair is selected randomly. Also, we note that while our methods does slightly worse, it is important to remember that the other methods assume known value of the parameters, whereas LPPM learns the parameters from the data.

\begin{figure}[!t]
\centering
\subfigure[]{
    \includegraphics[width=0.68\columnwidth]{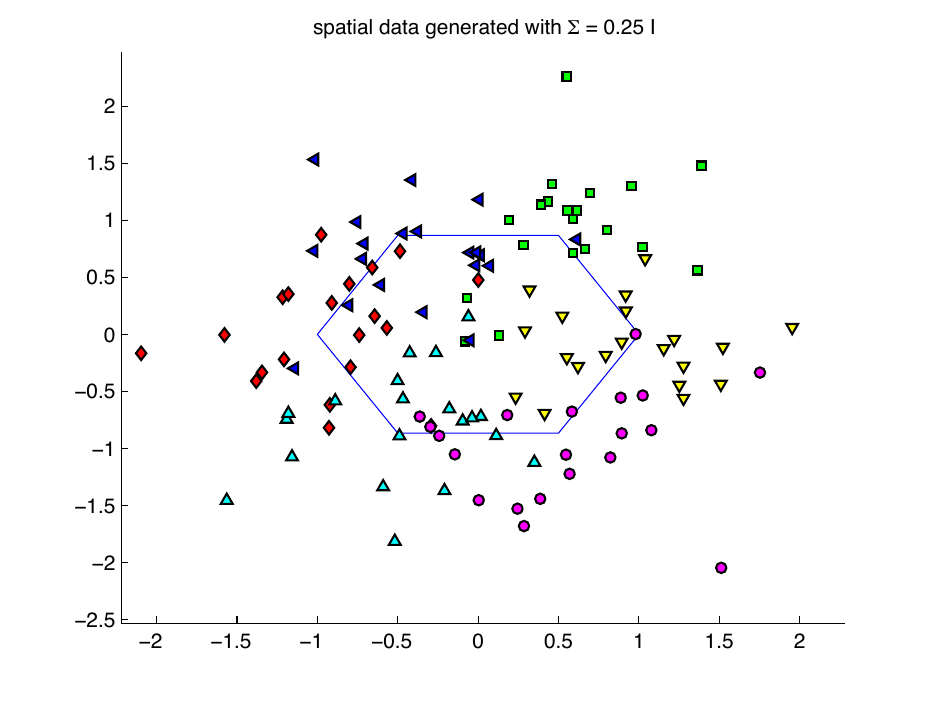}
}
\subfigure[]{
    \includegraphics[width=0.68\columnwidth]{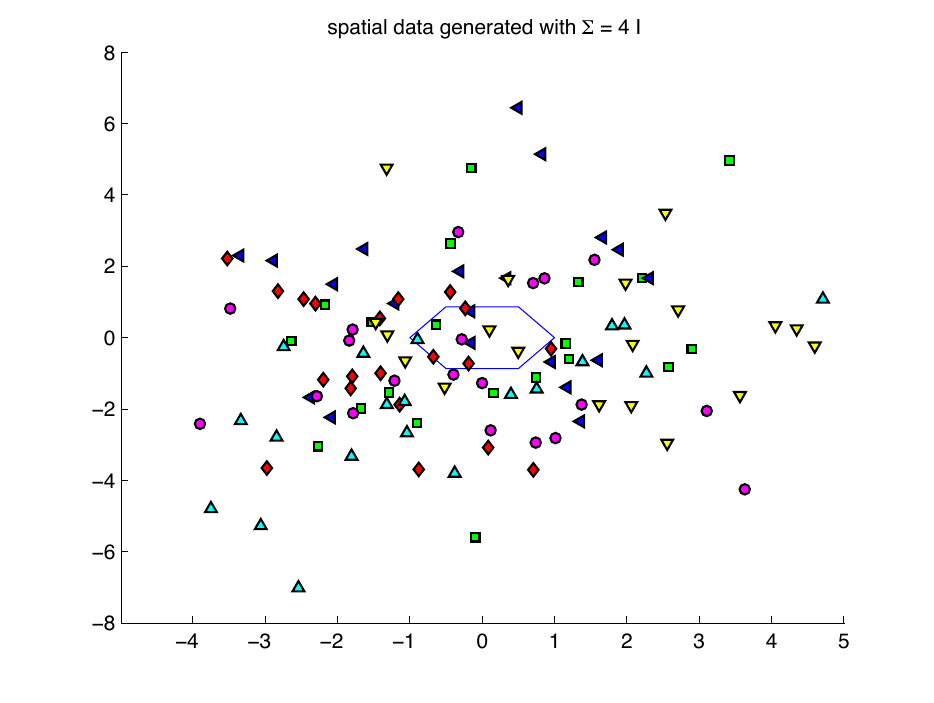}
}
\caption{The spatial data generated varying the covariance matrix from $0.25 \mathbf{I}$ (a) to $4 \mathbf{I}$ (b). Each color and symbol represents the pairs. 6 centers are on the vertex point of hexagon with side length 1, while the covariance matrix is being controlled}
\label{fig:hexagon}
\end{figure}

In the next set of experiments we  examine the relative importance of spatial and temporal parts by comparing three variants of our algorithm that use  $1.$ Temporal only data, $2.$ Spatial-only data, and 3. Combined spatial and temporal data. For the  spatial component of the data, we use six multivariate normal distributions with the center of each on the vertex point of hexagon (for all 6 pairs). Here we use simple Gaussian for each pair for the spatial process. As in Figure~\ref{fig:hexagon}, we fix the side length of the hexagon to 1, and analyzed how varying the width of the normal distribution affects the overall performance. Specifically, we varied the covariance matrix $\Sigma$ from $0.25 \mathbf{I}$ to $4  \mathbf{I}$. Again, the results are averaged  over 100 runs.  The accuracy was computed by averaging the number of correct estimates divided by the number of unknown incidents.

\begin{figure}[!t]
\centering
\subfigure[]{
    \includegraphics[width=0.68\columnwidth]{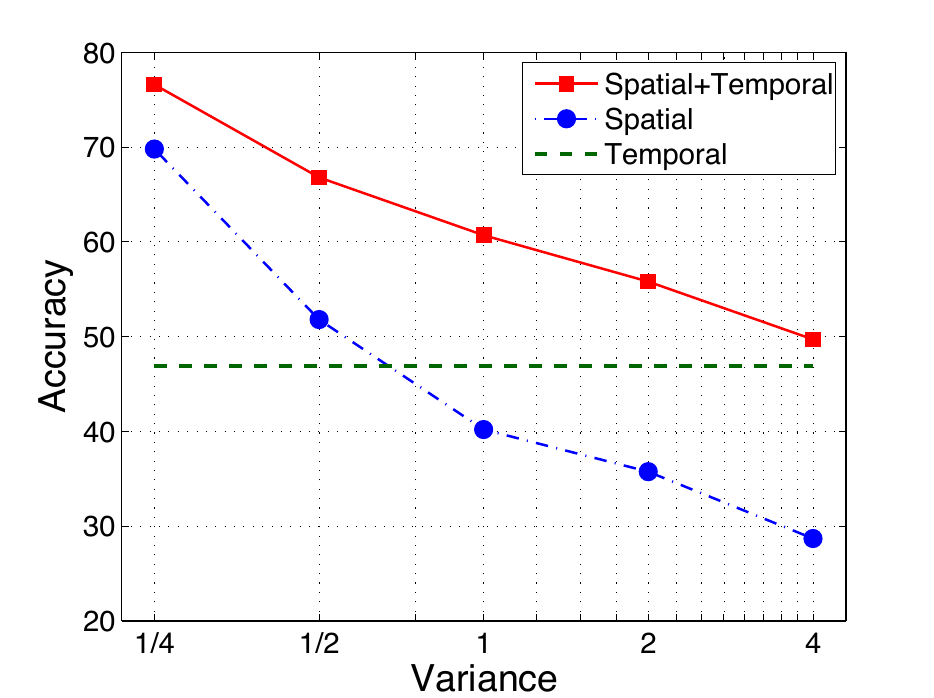}
}
\subfigure[]{
    \includegraphics[width=0.68\columnwidth]{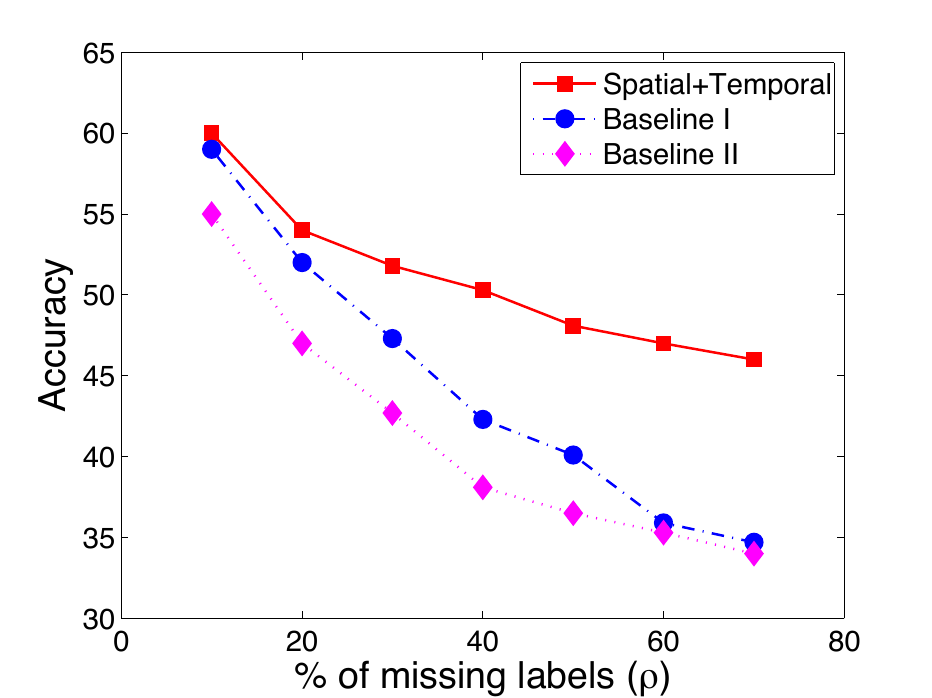}
}
\caption{(a) Accuracy of inference using spatial data only, temporal data only, and spatial-temporal data, for different settings of the standard deviation of the spatial Gaussian model. The results are averaged over 100 trials; (b) Average accuracy (over 20 trials) plotted against the percentage of missing labels. Spatial data was generated based on Gaussian with standard deviation 1.}
\label{fig:accuracy}
\end{figure}
As expected, the relative importance of the spatial information decreases when increasing $\sigma$. In the limit when $\sigma$ is very large, location of an event  does not contain any useful information about the participants, so that the accuracy based on spatial information only should converge to the random baseline $1/6$. On the other hand, for small values of $\Sigma$, the spatial information helps to increase accuracy.

In the last set of experiments with synthetic data, we examine the performance of LPPM  by varying the fraction of unknown incident labels. We compare the performance of LPPM to two baseline methods.
\begin{itemize}
\item Baseline I (B1): This method uses self-exciting Hawkes process model using {\em labeled} data only. We perform MLE to estimate the model parameters of Hawkes process by only considering the events that are labeled. In other words, we discard the events which misses the label: the information on the pairs.
\item Baseline II (B2): This method uses homogenous Poisson process model with constant intensity using {\em both}  labeled and unlabeled data.  This method is similar to our method except that the temporal process is based on Poisson process. This can be treated as a special case of Hawkes process with $\beta=0$.
\end{itemize}
We note that both LPPM and the baseline methods use the spatial component, so any differences in their performance should come from the temporal part of the model only.

The results of our comparative studies are shown in Figure~\ref{fig:accuracy}. It can be seen that LPPM outperforms both baselines by a significant margin, which increases as the data becomes more noisy.  Thus, LPPM is a much better choice when the amount of missing information is significant. The result also reflects that learning model parameters only with the labeled data is not sufficient for inferring missing labels.

\section{Experiments with real--world data}
\label{sec:real-world}
In this section we report on our experiments using two distinctly different real-world datasets. The first dataset describes gang-rivalry networks in  the Hollenbeck police division of Los Angeles~\cite{Tita},  and the second dataset is from a popular location-based social networking service {\em Gowalla}~\cite{Gowalla}. The rest of the section is organized as follows: We next describe both datasets; Then we conduct experiments on identity-inference problems in Section~\ref{sub:inferring}. Finally, we evaluate LPPM for event prediction problem in Section~\ref{subsub:prediction}

\subsection{Data description}
\label{descriptdata}

\begin{figure}[ht]
\vskip -0.1in
\centering
\subfigure[]{
    \includegraphics[width=0.7\columnwidth]{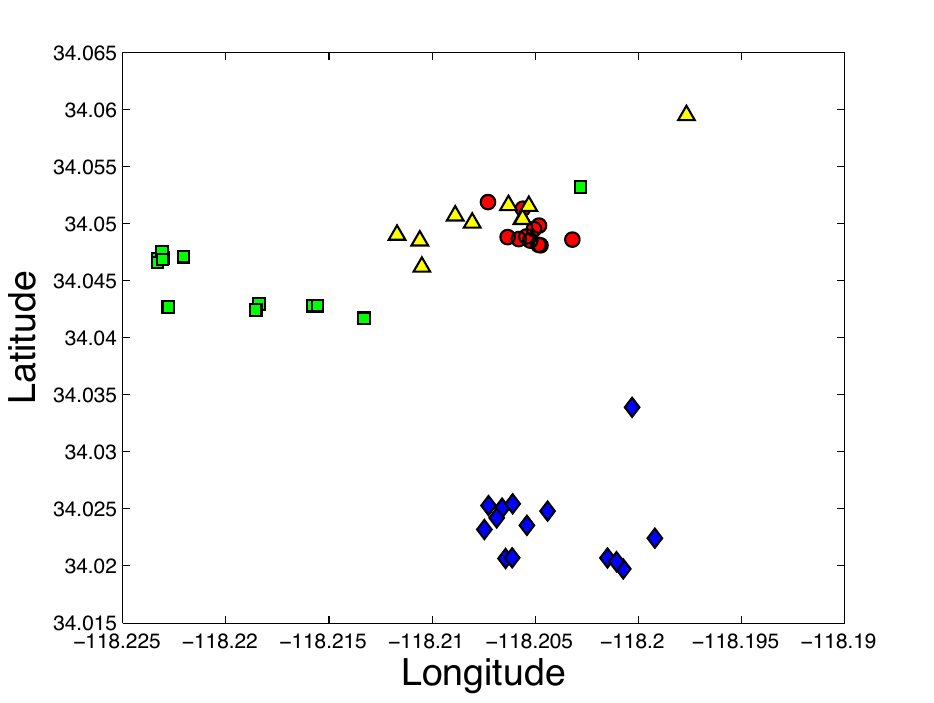}
}
\subfigure[]{
    \includegraphics[width=0.7\columnwidth]{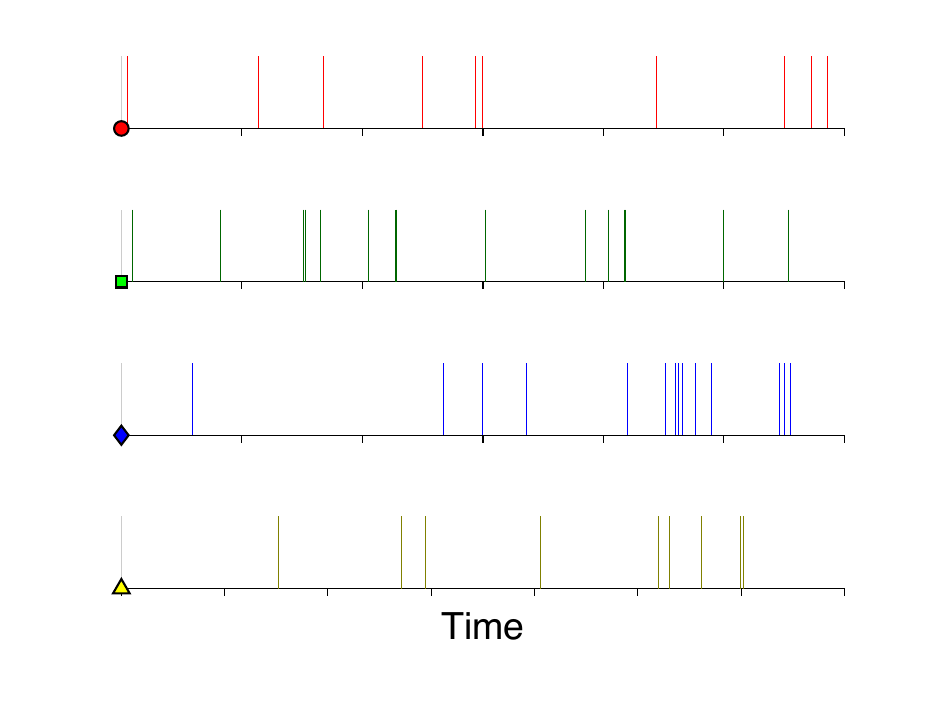}
}
\caption{Spatial (a) and temporal (b) description of the events involving four active gang rivalries. Different colors represent different pairs. In (b) each spike represents the time of the event.}
\label{fig:3relations}
\end{figure}

\paragraph{\textbf{LAPD dataset}} Hollenbeck is a 15.2 square mile (39.4 km2) policing division of the Los Angeles Police Department (LAPD), located on the eastern edge of the City of Los Angeles, with approximately 220,000 residents.  Overall, 31 active criminal street gangs were identified  in Hollenbeck between 1999-2002~\cite{Tita}. These gangs formed at least 40 unique rivalries, which are responsible for the vast majority of violent exchanges observed between gangs.  Between November 14, 1999 and September 28, 2002 (1049 days), there were 1208 violent crimes attributed to criminal street gangs in the area.  Of these, 1132 crimes explicitly identify the gang affiliation of the suspect, victim, or both.  The remaining events include crimes such as `shots fired' which are known to be gang related, but the intended victim and suspect gang is not clear.  For each violent crime, the  collected information includes the street address where the crime occurred as well as the date and time of the event~\cite{Tita}, allowing examination of the spatial-temporal dynamics of gang violence. In Figure~\ref{fig:3relations} we show temporal and spatial distribution of interactions between three most active gangs. For this dataset, we found that each pair is characterized by a simple Gaussian. This could be treated as a special case of GMM with $C=1$.

\vspace*{5pt}
\paragraph{\textbf{Gowalla dataset}} Gowalla is a location-based social networking website where users share their locations by checking-in \cite{Gowalla}.
We used the top 20 nodes who actively check-in to places. The network consists of 196,591 nodes and 950,327 undirected edges. 6,442,890 check-ins of these users were gathered from Feb. 2009 - Oct. 2010. Each check-in not only has its latitude and longitude coordinates but also has a given location ID provided by Gowalla. The location ID is very useful in that it enables us to verify the co-occurrence of a pair at a given location even though the location of latitude and longitude has some error or has a multi-story building at the given coordinates. Gowalla also has a list of friends, where the edge between them is undirected. We looked into every check-in of the friends of 20 nodes and assumed they have interacted each other if the check-in of the two at same location was within 10,000 seconds. The venues of popular places such as airport and stations has been removed to rule out the unexpected coincidence between users. Out of 20 active nodes, we were able to collect 3 groups: one from Stockholm, Tokyo, and San Francisco.

\subsection{Inferring event participants}
\label{sub:inferring}
As we mentioned earlier, most social network data is noisy and incomplete with missing information about nodes and/or interactions. In this section, we consider a scenario where one has the timing and location of interaction events, but only partial information about event participants. A specific real-world problem described within this scenario is inter-gang violence, where one has a record of reported violent inter-gang events, but where either the perpetrator gang, the victim gang, or both, are unknown. Thus, the problem is to infer the unknown participants based on available information. The naive solution would be to discard the missing data, learn the model parameters based on fully observed events only, and then use the learned model for inferring participants of partially labeled events. However, below we show that the naive approach is sub-optimal. Instead, by taking into account missing data via the expectation-maximization framework, one achieves better  accuracy in the participant identification task.

\subsubsection{Experiments with LAPD dataset}
As described above, the LAPD dataset contains the time stamp and the location of incidents between pairs of gangs.  Approximately $31\%$ of the records contain information about both participants in the event. Furthermore,  $62\%$ of the records contain information about one of the participants, but not the other. Finally, $7\%$ do not have any information about the participants. For better understanding of gang-rivalries, it is important to recover missing information on those $~70\%$ of the whole data. Since this research is not the studies of the actual rivalries in Hollenbeck but to verify how well our algorithm performs on inference, in the experiments below, we discard the latter portion of the data.This way we could validate our inference and by comparing it with actual given label. In the remaining data,  we focused on $31$ active gangs which were involved in at least  $4$ incidents within the time period. Furthermore, out of all possible pairs, we use 40 pairs which had more than one reported  incident between each other.

In the first set of experiments, we focused on the portion of the data that contains information about both participants. We randomly select a fraction $\rho$ of the incidents, and then hide the identity of the participants for those incidents. Next, we use LPPM  to see how well it can reconstruct the hidden identities by varying $\rho$. We compared the results to the same two baseline methods outlined in Section~\ref{sec:results-synthetic}. In addition, we add another baseline that uses all existing labels to learn a spatial-only model. The accuracy is defined as the fraction of events for which the algorithm correctly recovers {\bf both} participants. The results were averaged over $20$ different runs. The center of clusters were initialized with the mean location of labeled data.

\begin{figure}[!t]
\vskip 0.2in
\begin{center}
{\includegraphics[width=0.74\columnwidth]{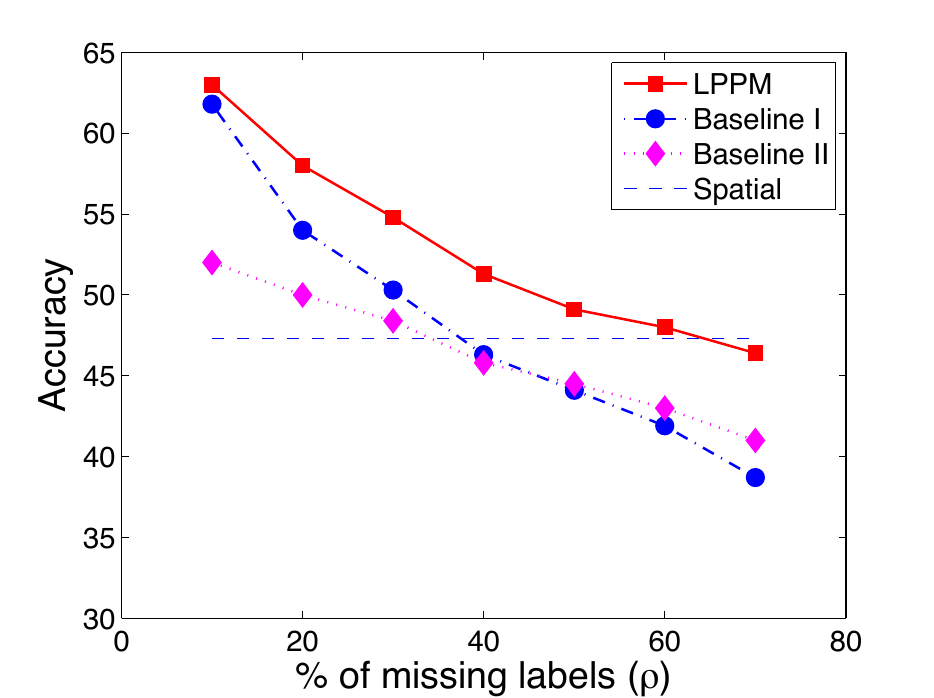}}
\caption{Average accuracy for varying fraction of missing labels. Baseline I and Baseline II are defined in Section~\ref{sec:results-synthetic}. The horizontal line corresponds to inference using spatial data only.}
\label{fig:ratio_vs_accuracy}
\end{center}
\end{figure}

Figure~\ref{fig:ratio_vs_accuracy} demonstrates our results.  One can see that the LPPM does consistently better than B1 and B2. For only $10\%$ of missing label information,  the accuracy of  LPPM and B1 are fairly close. This is to be expected, since for vanishing $\rho$ those algorithms become identical -- they learn the same model using the same data. However, LPPM performs much better than B1 when $\rho$ increases. Another interesting observation is that B2 performs better than B1 when $\rho$ is sufficiently large. This suggests that for large $\rho$ it is better to use a {\em simpler} (and presumably wrong) model using both missing and labelled data, than learn a more elaborate model using labelled data only.

We also note LPPM does better than the spatial-only baseline even when half of the events are hidden. This is significant since the spatial model uses all the label information that is not available to LPPM. Although the  spatial model performs better when $\rho$ increases further, LPPM remains very competitive even when $70\%$ of the events are hidden, which is the same condition (i.e., fraction of unknown) of LAPD gang related crime data.

\subsubsection{Experiments with Gowalla dataset}

Next, we perform experiments on the participant-inference task using the Gowalla data. Note that while the participant information is generally available in this data, it still provides an interesting benchmark for validating LPPM.

Out of 20 most active users in Gowalla network, we focus on three  users that have high interaction frequency with their friends.~\footnote{Recall that for this dataset, an interaction between two users is determined by near-simultaneous check-ins; see the description of the dataset}  Coincidently, three users were from different city (Tokyo, Stockholm, and San Francisco). We found that some of the check-in locations were repeated by the same pairs. Strictly speaking, this suggests that the spatial component is not a point process. However, this detail has little bearing on our model, as the spatial interactions can still be modeled via the Gaussian mixture model.

Spatial analysis of the dataset reveals that the interaction are multi-modal in the sense that the same pair of users interact at different locations. This is different from the crime dataset, and necessitates  using more than one component for the spatial mixture model. In the experiments, we used 4 components of GMM for two of the pairs (Stockholm and San Francisco), and three components for the other pairs (Tokyo).

\begin{figure}[ht]
\begin{center}
    \includegraphics[width=0.7\columnwidth]{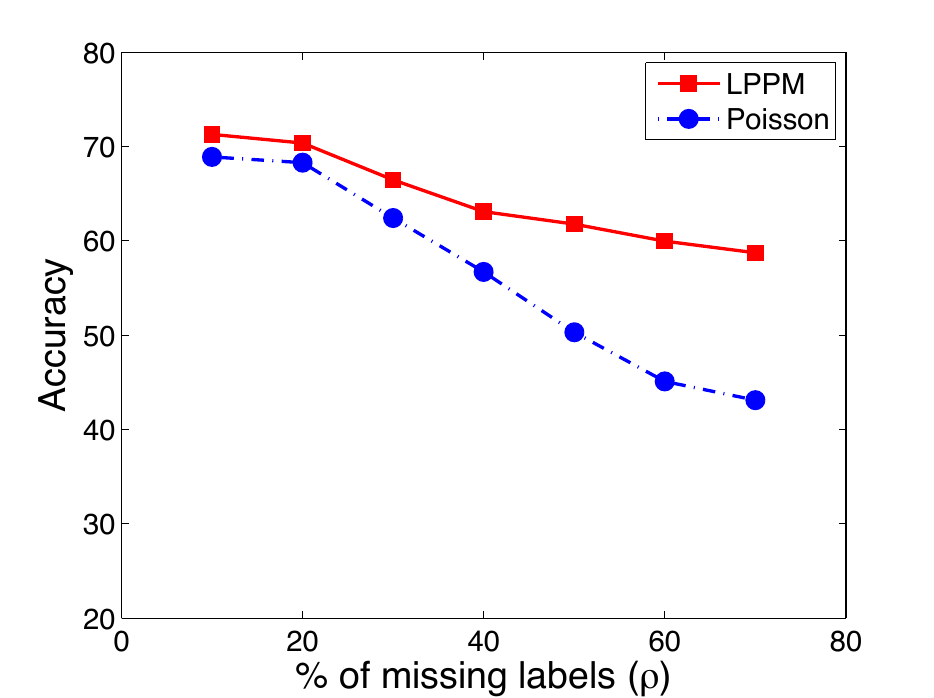}
    \label{SF}
\caption{Average accuracy of participant-inference task for the user in San Francisco. The fraction of missing labels is varied between 10\% and 70\%.}
\label{fig:mc_compare}
\end{center}
\vskip -0.1in
\end{figure}

The results of the experiments are shown in Figure~\ref{fig:mc_compare}. Due to limited space, we present the result of simulation using users in San Francisco. Since the two baseline methods perform similarly, here we show the comparison only with B2, which learns a homogenous Poisson point process model using both labeled and unlabeled data. Again, the results suggest that LPPM is consistently better than the baseline for all of the pairs. The gap between LPPM and the baseline is not significant as before which is mainly due to the active pairs which dominates the interactions. When there are dominant active pairs, Poisson process could distinguish the users by comparing the rate between the pairs. Moreover, there were some active pairs which have checked into the exact same location repeatedly leading to higher accuracy.

\subsection{Event prediction with LPPM}
\label{subsub:prediction}
LPPM can be used not only for inferring missing information but also predicting future events, which can be potentially useful for many applications. For instance, in the context of proactive policing, the predictions can be used to  anticipate the participants/timing/location of the next event, and properly assign resources for patrol, etc. Related to friendship network, one can predict the spatial-temporal movement patterns by predicting the hot clusters involving given pairs. This kind of prediction can be also very useful in epidemiology, i.e., by predicting  diffusion patterns of an infectious disease.

  In this section, we use learned LPPM models for two different prediction tasks: \\(1) Predicting the timing of the next interaction event.  \\(2) Predicting the pair that will have the next interaction.

Let us first discuss the timing prediction problem. Given the history of events up to  the $k$-th event, our goal is to predict the timing of the $(k+1)$-th event. Note that the prediction can be either pair-specific, or across all pairs. Here we select the latter option.

The estimated waiting time until the next incident is given by
\begin{eqnarray}
\int_0^L  t \lambda_\mathcal{S} (t) \exp({-\int_0^t \lambda_\mathcal{S} (\tau) d\tau})  dt
\end{eqnarray}
where $L$ is fairly a large number, and $\lambda_\mathcal{S} (t)=\ssum_{(ij)}\lambda_{ij}(t)$ is the sum of the conditional intensity function across all the pairs. Below we compare the prediction performance of LPPM with the B2 defined in Section~\ref{sec:results-synthetic},  which employs homogenous  Poisson processes. According to this baseline, the expected waiting time to the next event is simply $1 \over \ssum_{(ij)}\lambda_{ij}^*$ ($\lambda_{ij}(t)\equiv \lambda_{ij}^*$), where $\lambda_{ij}^*$ is the time-independent intensity for the pair $(i,j)$.

The prediction accuracy is measured using the {\em mean absolute percentage error} (MAPE) score, which measure the relative error of the predicted waiting time: ${\text{MAPE}}=|{A_n-F_n \over A_n }|$, where $A_n$ is actual waiting time until the next incident, and $F_n$ is our predicted value. Note that more accurate prediction corresponds to lower MAPE score, ${\text{MAPE}}=0$ for perfect prediction.

We measure the MAPE score for LPPM prediction on the LAPD and Gowalla datasets. For the former, we use LPPM to predict the timing of the last 50 incidents among top 40 pairs. As for the latter dataset,  we focus on only one of the users (in Tokyo), and use the last 10 events (out of 40 total) for prediction. For both datasets,  LPPM provides significantly more accurate prediction than the baseline for most of the incidents. LAPD dataset had \textbf{2.7502} for LPPM compared to \textbf{11.0434} for B2; Gowalla dataset had \textbf{1.2236} for LPPM compared to \textbf{5.9350} for B2. A possible explanation of the poor performance of  the Poisson model is that it fails to accurately  predict the timing of highly correlated events that are clustered in time, whereas LPPM is able to capture such correlations. When the next event is highly influenced by the previous event, Poisson model is limited in that it considers the triggered event as a random event.


For the prediction task  (2), we used LPPM  to find the conditional intensity of interactions between different pairs based on all the events up to event $k$ that happens at time $t_k$. We then predict that the pair with the highest conditional intensity to have an interaction event at a time $t>t_{k}$, assuming that no other interaction has taken place in time interval $[t_k,t]$. Note that the homogeneous Poisson process model (Baseline II) simply  selects the pair that has been the most active in the past. For this particular task, we also use another prediction method (Baseline III) which predicts that the pair that had the last event will also participate in the follow-up event.  In addition to the top pair, we also predict the second  and third best predictions. We performed experiments with the crime dataset,  for which 14 incidents out of 100 were predicted correctly by LPPM. Baseline II correctly predicted only  8 incidents, whereas Baseline III did considerably better with 13 correct predictions. Furthermore, LPPM outperforms both methods in predicting top 2 and top 3 users, as shown in Table~\ref{table:inference_comparison}.
\begin{table}[!h]
\caption{Prediction accuracy of top-K choices for K=1,2,3.}
\label{table:inference_comparison}
\vskip 0.15in
\begin{center}
\begin{small}
\begin{sc}
\begin{tabular}{lcccr}
\hline
Method & Baseline II & Basline III & LPPM \\
\hline
Top 1    &8\%& 13\% & 14\% \\
Top 2   &16\%& 20\% & 26\%\\
Top 3    & 23\% &22\% & 37\% \\
\hline
\end{tabular}
\end{sc}
\end{small}
\end{center}
\vskip -0.15in
\end{table}

\section{Conclusion}
\label{sec:conclusion}
We suggested a latent point process model to describe spatial-temporal interaction networks. In contrast to existing continuous time models of temporal networks, here we assume that interactions along the network links are only partially observable. We describe an efficient variational EM approach for learning and inference with such models, and demonstrated a good performance in our experiments with both synthetic and real-world data.

We note that while our work was motivated by modeling spatial-temporal interaction networks, the latent point process suggested here is much more general and can be used for modeling scenarios where one deals with latent mixture of arbitrary point processes. For instance, LPPM can be generalized to describe geographically distributed sequence of arbitrary events of multiple pairs even with the events which misses the pair information.

There are several ways to generalize the model further. For instance we have assumed a homogenous background rate, whereas in certain scenarios one might need to introduce cyclic activity patterns. Furthermore, the assumption that the process intensity is factorized into temporal and spatial components might not work well for certain types of processes, where the location component might depend on the event time.

\section*{Acknowledgments}
This research was supported in part by the  US AFOSR MURI grant FA9550-10-1-0569, US DTRA grant HDTRA1-10-1-0086, and  DARPA grant No. W911NF-12-1-0034.

\appendix

\gdef\thesection{ \Alph{section}}
\section{Variational E-step}
In the variational E-step, we maximize $\mathcal{L}_{\Phi}$  over the variational parameters. Note that the variational parameters shoud satisfy the normalization constraint $\sum_{i<j} \phi_p^{ij} = 1$. By introdcuing Lagrange multipliers $\gamma_p$ to enforce those constraints, and taking the derivative of Equation~\ref{eq:lowerbound2} with respect to the variational parameters yields 
 \begin{eqnarray}
0&=& {\partial \over \partial \phi_p^{ij}} E_Q\bigl[ \sum_k  z_k^{ij} \log  [\lambda_{ij} (t_k) ] {-\Lambda_{ij}^T } \bigr] \nonumber \\
 &+&   \log  [r_{ij} (\bx_p) ]  \nonumber \\
&-&    \log   {\phi}_k^{ij} -1 +\gamma_{p}
 \label{eq:difflowerbound}
 \end{eqnarray}
 Solving the constrained optimization problem with Lagrange multipliers, we have the update equation for variational parameter $\phi_p^{ij}$ as below:
\begin{equation}
\label{eqn_phi}
\phi_{p}^{ij} = {1\over C_p} \exp\biggl \{ \frac {\partial}{\partial \phi_{p}^{ij}} E_Q \biggl [{\sum_{k} z_{k}^{ij} \log \lambda_{ij}(t_k)  } -\Lambda_{ij}^T
   \biggr ] \biggr\} [r_{ij} (\bx_p) ]
\end{equation}
where the Lagrange multiplier has been absorbed in the normalization constant $C_p$.

For the evaluating the derivative of the expectation of $\log \lambda_{ij}(t_k)$ with respect to $\phi_p^{ij}$ in the above equation, we separate into two cases when $k>p$ and $k=p$. Before expressing the derivatives for two cases, we introduce a new function for a simpler expression.
\begin{equation}
\mathcal{M}_{ij}(\mathcal{Z}_k) = \prod _{l=1}^{k}  {\phi_{l}^{ij}} ^{z_{l}^{ij}} {(1- \phi_{l}^{ij} )}^{(1-{z_{l}^{ij}})}
\end{equation}
which is a joint probability of given scenario from the beginning up to event $k$.

First, for the case when $k=p$, we have
\begin{equation}
\frac{\partial}{\partial \phi_{p}^{ij}} E_Q[\log \lambda_{ij}(t_p)] =
 \sum _{\mathcal{Z}_{p-1}}  \mathcal{M}_{ij}(\mathcal{Z}_{p-1})      
 \log \biggl [\mu_{ij}  + \sum _{l=1}^{p-1} z_{l}^{ij} g_{ij}(t_p-t_l)  \biggr ]
\label{eq:15}
 \end{equation}
 In the right hand side of  Equation~\ref{eq:15}, the sum is over all the possible configurations of the latent variables up to the event $p-1$, $Z_{k=1}^{p-1}$.
Similarly, we can derive the derivative with respect to $\phi_{p}^{ij}$ for the terms with $k>p$.
For steps when $k$ is greater than $p$,
\begin{eqnarray}
\frac{\partial}{\partial \phi_{p}^{ij}} E_Q[\log \lambda_{ij}(t_k)] =
 \sum _{{ \tilde{\mathcal{Z}}_{k-1}^p }}  \phi_k^{ij}  \tilde{\mathcal{M}}_{ij}^p(\tilde{\mathcal{Z}}_{k-1}^p) ~~~~~~~~~~~~~~~~~~~~~~~~~~~~~~~~~~~~~~~~~~~~~~~~~~~~~~~~~~~~~~~~~~~~~~~~~~~~~~~~~~~~
\\ \times
  \log \biggl [ {\mu_{ij}  + \sum _{l=1,l\neq p }^{k-1} z_{l}^{ij} g_{ij}(t_k-t_l) + g_{ij}(t_k-t_p)  \over \mu_{ij}  + \sum _{\substack{{l=1}\\ {l \neq p}}}^{k-1} z_{l}^{ij} g_{ij}(t_k-t_l) } \biggr]   \nonumber
\label{eq:updatephi}
\end{eqnarray}
where we have defined $\tilde{\mathcal{Z}}_{k}^p$ as ${\mathcal{Z}}_{k}$ excluding $z_p$ with $\tilde{\mathcal{M}}_{ij}^p(\cdot)$  following the same logic. The numerator term in the logarithm above comes from when pair $i$ and $j$ trigger the $k$-th event on the $p$-th event, while the denominator term comes from when they did not.

Finally for the derivative of expectation of $\Lambda_{ij}^T$ in Equation~\ref{eqn_phi}, we use
\begin{equation}
\frac{\partial}{\partial \phi_{p}^{ij}} E_Q[-\Lambda_{ij}^T] = - \beta_{ij} \{1-\exp(\omega_{ij} (T-t_p) ) \}
\label{eqn_Lambda}
\end{equation}

By combining Equation~\ref{eqn_phi}  -- \ref{eqn_Lambda}, we obtain an iterative scheme for finding the variational parameters of the form
\begin{equation}
\phi_{p}^{ij} = f(\{ \phi_{p}^{ij} \}_{k=1;k\neq p}^n; \Theta)
\label{eq:phi_func}
\end{equation}
 The above iterations are used until the convergence of all the variational parameters.

\section{Variational M-step}
The M-step in the EM algorithm computes the parameters by maximizing the expected log-likelihood found in the E-step. The model parameters consists of spatial parameters and temporal parameters. We first look into the update equations of spatial parameters.
For some cases, when the spatial pattern is distinct over pairs, we use single Gaussian for each pair, and the update equations are as below (i.e., the mean and the variance of Gaussian distribution):
\begin{eqnarray}
\mathbf{m}_{ij} &\leftarrow&{ \sum_k  \phi_{k} ^{ij} \bx_k \over  \sum_k \phi_{k} ^{ij} } \\
\mathbf{\sigma}_{ij,lat}^2 &\leftarrow&{ \sum_k  \phi_{k} ^{ij} (\mathbf{x}_{k,lat} - \mathbf{m}_{ij,lat} )^2  \over  \sum_k \phi_{k} ^{ij} }  \\
\mathbf{\sigma}_{ij,long}^2 &\leftarrow&{ \sum_k  \phi_{k} ^{ij} (\mathbf{x}_{k,long} - \mathbf{m}_{ij, long} )^2 \over  \sum_k \phi_{k} ^{ij} }
\end{eqnarray}
When using a Gaussian mixture model, the weight vector of the mixture model for each pair is updated respectively.
\begin{equation}
w_{ij}^c \leftarrow { \sum_k \phi_k^{ij} {{\mathcal{N}(\mathbf{x}_k | \mathbf{m}_{ij}^c, \Sigma_{ij}^c )}\over {\sum_{p=1}^C \mathcal{N}(\mathbf{x}_k | \mathbf{m}_{ij}^p, \Sigma_{ij}^p )} } \over \sum_k \phi_k^{ij}}
\end{equation}

The re-estimation of the temporal parameters are more involved. For instance, to estimate $\mu_{ij}$, we nullify the derivative of the likelihood with respect to $\mu_{ij}$, $\frac{\partial \mathcal{L}_{\Phi}}{ \partial \mu_{ij}} =0$, which yields
\begin{equation}
\mu_{ij} \leftarrow {\sum_k \sum _{\mathcal{Z}_{k-1}}    \phi_{k}^{ij} {  \mu_{ij} \mathcal{M}_{ij} (\mathcal{Z}_{k-1})  \over \mu_{ij}  + \sum _{l=1}^{k-1} z_{l}^{ij} g_{ij}(t_k - t_l)    }  \over T}
\end{equation}
%
Similarly, for re--estimation of $\beta_{ij}$, we present the derivative as below:
\begin{equation}
\beta_{ij} \leftarrow  {\sum_k \sum _{\mathcal{Z}_{k-1}}    \phi_{k}^{ij} {   \mathcal{M}_{ij} (\mathcal{Z}_{k-1}) \sum _{l=1}^{k-1} z_{l}^{ij} g_{ij}(t_k - t_l)   \over \mu_{ij}  + \sum _{l=1}^{k-1} z_{l}^{ij} g_{ij}(t_k - t_l)    }  \over \sum_k \phi_k^{ij}   \int_{0}^{T-t_k}   \omega_{ij} e^{-\omega_{ij} \tau}   d\tau  }
\end{equation}
Finally, for $\omega_{ij}$, we obtain
\begin{eqnarray}
\label{eq:omega}
\\
\sum_k \phi_k^{ij} \biggl[ \sum _{\mathcal{Z}_{k-1}}   \bigl[  { \bigl( \sum \limits_{l=1}^{k-1} z_{l}^{ij} (1-\omega_{ij}(t_k-t_l) )g_{ij}(t_k-t_l) \bigr)    \over \mu_{ij}  + \sum _{l=1}^{k-1} z_{k}^{ij} g_{ij}(t_k - t_l)    }  \nonumber \\
\times  \mathcal{M}_{ij}(\mathcal{Z}_{k-1}) \bigr]
 -  \beta_{ij} (t_k - T)  e^{-\omega_{ij}( T-t_k) }       \biggr]
=  0\nonumber
\end{eqnarray}
where
\begin {equation}
\label{eqn:selfexciting}
g_{ij}(t-t_p) = \beta_{ij} \omega_{ij} \exp\{-\omega_{ij} (t-t_p)\}
\end{equation}
Unfortunately, the resulting equations do not allow closed form solutions, so they have to be solved using numerical methods, such as the  Newton's  method employed here. We can also have closed form of update equation of $\omega_{ij}$ by approximating the second term to zero in Equation~\ref{eq:omega}. When $\omega_{ij}T$ is fairly large compared to $\omega_{ij}t_k$, we can ignore the second term, and have closed form as below:
\begin{equation}
\label{eq:omega_closed}
\omega_{ij} \leftarrow  {\sum_k \sum _{\mathcal{Z}_{k-1}}    \phi_{k}^{ij} {   \mathcal{M}_{ij} (\mathcal{Z}_{k-1}) \sum _{l=1}^{k-1} z_{l}^{ij} g_{ij}(t_k - t_l)   \over \mu_{ij}  + \sum _{l=1}^{k-1} z_{l}^{ij} g_{ij}(t_k - t_l)    }  \over \sum_k \sum _{\mathcal{Z}_{k-1}}    \phi_{k}^{ij} {   \mathcal{M}_{ij} (\mathcal{Z}_{k-1}) \sum _{l=1}^{k-1} z_{l}^{ij} (t_k-t_l)g_{ij}(t_k - t_l)   \over \mu_{ij}  + \sum _{l=1}^{k-1} z_{l}^{ij} g_{ij}(t_k - t_l)    }  }
\end{equation}

The following remark is due: the update equations for both the variational parameters and the model parameters involve summation over the all possible configurations of the latent variables. This sum might become prohibitively extensive for long history windows. However, due to the exponential decay of the self-excitation term, events too far in the past have negligible impact on future events. This observation justifies  limiting the summation to a window, i.e.,   $\lambda_{ij}(t_k|\H_{t_k}) \approx \lambda_{ij}(t_k| \{h_l\}_{l=k-d}^k) $, which discards  events that are far in the past. In the results, we use this truncation to speed up the inference process.

\medskip
Received December 2012; revised April 2013.
\medskip

\end{document}